\documentclass[preprint,12pt]{elsarticle}

\usepackage{amssymb}
\usepackage{url}

\begin{document}

\begin{frontmatter}



\title{First (calibration) experiment using proton beam from FRENA at SINP}


\author[1,2]{C. Basu}
\ead{chinmay.sinp@gmail.com}
\author[3,2]{K. Banerjee}
\author[3,2]{T. K. Ghosh}
\author[3,2]{G. Mukherjee}
\author[3,2]{C. Bhattacharya}
\author[4]{Shraddha S Desai}
\author[5]{R. Shil}
\author[3]{A. K. Saha}
\author[3]{J. K. Meena}
\author[1,2]{T. Bar}
\author[1,2]{D. Basak}
\author[1,2]{L.K. Sahoo}
\author[1,2]{S. Saha}
\author[1]{C. Marick}
\author[1]{D. Das}
\author[1]{D. Das}
\author[1]{D. Das}
\author[1]{M. Kujur}
\author[1]{S. Roy}
\author[1]{S. S. Basu}
\author[1]{U. Gond}
\author[1]{A. Saha}
\author[1]{A. Das}
\author[1]{M. Samanta}
\author[1]{P. Saha}
\author[1]{S. K. Karan}

\address[1]{Saha Institute of Nuclear Physics, 1/AF Bidhannagar, Saltlake, Kolkata-700064, India}
\address[2]{Homi Bhabha National Institute, Anushaktinagar, Mumbai-400094, India}
\address[3]{Variable Energy Cyclotron Centre, 1/AF Bidhannagar, Saltlake, Kolkata-700064, India}
\address[4]{Bhabha Atomic Research Centre, Trombay, Mumbai-400085, India}
\address[5]{Siksha Bhavana, Visva-Bharati, Santiniketan-731235, India}

\begin{abstract}
This work presents the first calibration experiment of a 3 MV Tandetron accelerator, FRENA, performed in May 2022. The $^7$Li(p,n) reaction threshold was measured  to calibrate the terminal voltage measuring device. A LiF target of thickness 175 $\mu$g/cm$^2$ was used in the experiment. The measured threshold was 1872$\pm$2.7 keV, indicating 6$-$10 keV energy shift.
\end{abstract}



\end{frontmatter}


\section{Introduction}
Facility for Research in low energy Nuclear Astrophysics (FRENA), a high current low energy (0.2$-$3 MV) Tandetron accelerator facility has been commissioned in the Saha Institute of Nuclear Physics, Kolkata. The machine is presently in the trial period of AERB and is delivering proton beams at a maximum current of 100 e$\mu$A. FRENA is primarily dedicated for performing experiments related to nuclear astrophysics. Being a DC accelerator, its beam energy is directly proportional to the terminal voltage. Therefore, before the initiation of actual experiments it is necessary to properly calibrate the terminal voltage of the accelerator. This can be done by measuring threshold energy of cetrain reactions and nuclear resonances within the FRENA energy range~\cite{Rajta2018}. A series of threshold reactions and nuclear resonances are described in Ref.~\cite{Marion1966}. Accordingly, a set of known threshold and resonance reactions, such as (p,n), ($\alpha$,n), and (p,$\gamma$), ($\alpha$, $\gamma$) have been planned to be conducted in FRENA. As a first stage, the $^7$Li(p,n)$^7$Be reaction was studied, details of which are described here.

\begin{figure}[!h]
     \centering
    \includegraphics[scale=0.3]{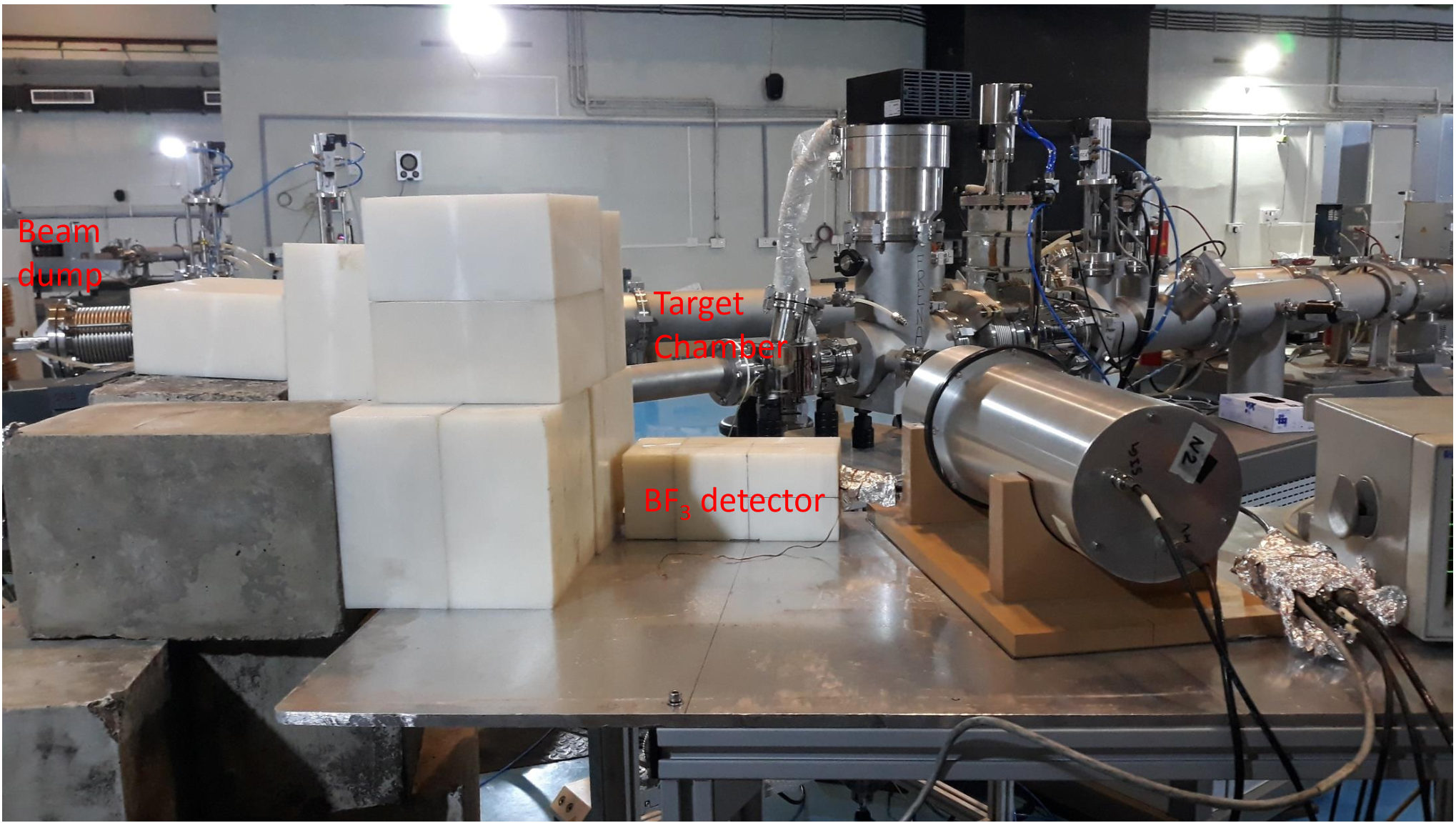}
\caption{Experimental setup for the $^7$Li(p,n)$^7$Be reaction.}
\label{fig:fig1}
\end{figure}
Experiment was set up at the PBA (Pulsed Beam Analysis) beamline of the FRENA accelerator. The mechanical arrangement in this beam line consists of a pneumatic gate valve which isolates the target chamber from the accelerator side to avoid accidental vacuum degradation. A 152 mm diameter stainless steel target chamber having wall thickness of 4 mm was coupled to the beam line for placing the targets through the bottom flange. A target ladder with three holders was used which contains; a 175 $\mu$g/cm$^2$ LiF on 130 $\mu$g/cm$^2$ Al backing for actual experiment, a thick Al$_2$O$_3$ plate for beam spot monitoring and one blank frame for background run respectively. A camera was coupled to the out of plane port at 150 $^{\circ}$ to monitor the beam spot. The chamber was followed by a 3KW Faraday cup to dump the beam and measure its current. A photograph of the experimental setup is shown in Fig. 1.
\begin{figure}[ht]
\centering
\includegraphics[scale=0.35]{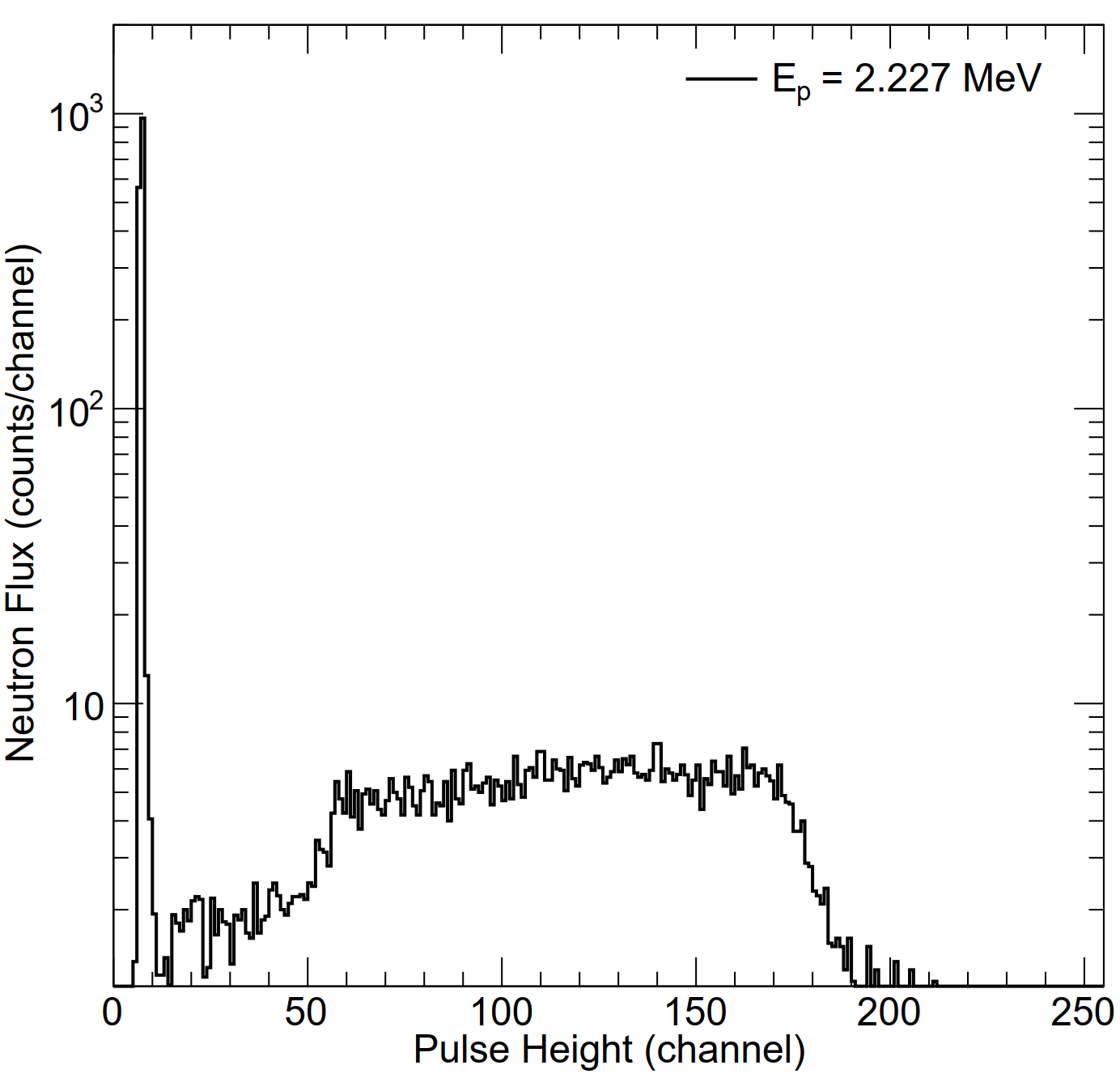}
\caption{Pulse height spectrum of the BF$_3$ detector.}
\label{fig:fig2}
\end{figure}
The experiment was performed using proton beam of energy ($E_p$) 1.78 MeV to 2.5 MeV with typical beam current of $\sim$100 nA. A cylindrical BF$_3$ detector of 10 mm diameter and 10 cm long with a sensitivity of 0.3 cps/nv was housed inside a high-density polyethylene moderator block to measure the emitted neutrons. The detector was kept at a distance of 1 meter from the target centre and normal to the beam direction to minimise background neutron contribution from the beam dump. For further reduction of neutrons/$\gamma$ background, beam dump was shielded with concrete, high density polyethylene and lead blocks.  Neutron pulse height spectra were measured for each proton energy. A typical pulse height spectrum obtained for $E_p$ = 2.227 MeV is shown in Fig. 2. The sharp peak near the zero channel is due to the $\gamma$ events and electronic noise whereas the broad distribution is due to the neutrons. Pulse height spectra were also measured using blank frame in the target position for each beam energy to get the background which is found to be negligibly small for the measured energy range.
Pulse height spectra obtained for each beam energy were integrated to obtain neutron flux for a given $E_p$. Fig. 3(a) shows the plot of neutron flux per beam charge vs. proton energy $E_p$. The figure shows two resonances at the at $E_p$ = 1.91 and 2.25 MeV. Energy dependence of the total neutron cross section just above threshold varies as($E_p$-$E_{th}$)$^{3/2}$~\cite{Brindhaban1994, Sorieul2021}, where $E_{th}$ is the threshold energy. Therefore, to fit the data using linear functions the 2/3$^{rd}$ power of the neutron flux per beam charge was plotted as a function of $E_p$ [Fig 3(b)]. Data points below and above the known threshold ($E_{th}$ =1.88 MeV) show two different slopes. Two different linear functions were used to fit the data points of these two groups which are shown with red and magenta lines. The intercept point of these two linear fits has been considered as the measured threshold of $^7$Li(p,n)$^7$Be reaction which is found to be 1872.5 $\pm$ 2.8 keV in this case.

\begin{figure}[h!]
\centering
\includegraphics[scale=0.4]{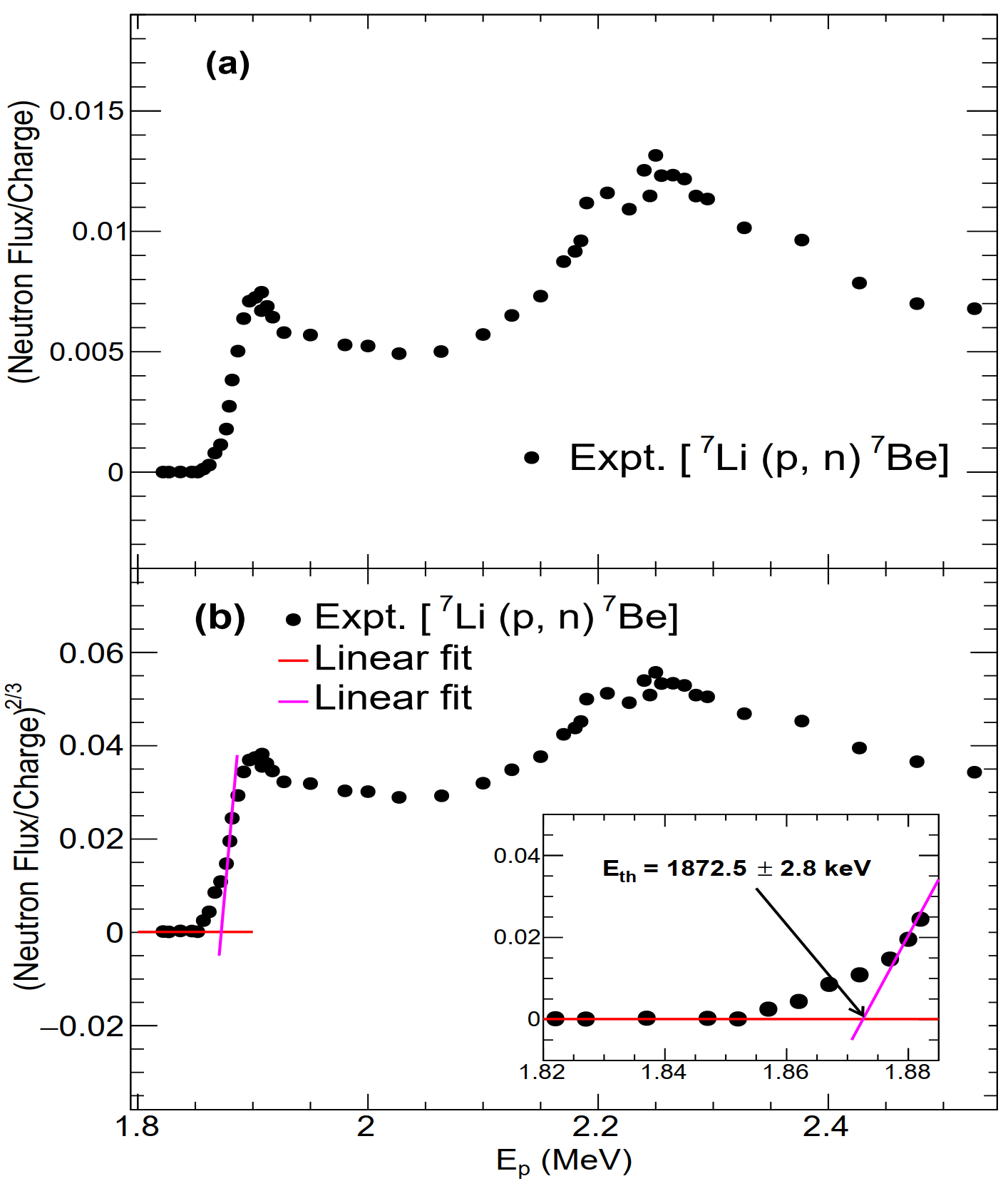}
\caption{(a) Neutron flux per beam charge as a function of proton energy $E_p$. (b) The 2/3$^{rd}$ power of neutron flux per beam charge as a function of $E_p$ along with two linear fits (see text).}
\label{fig:fig3}
\end{figure}

 Further experiment using other reactions has been planned which will be taken up to cover the entire range of the acclerator terminal voltage and a global calibration plot between measured threshold as a function of terminal voltage will be prepared.

\end{document}